\documentclass[aip,amsmath,amssymb,reprint]{revtex4-1}

\usepackage{graphicx}
\usepackage{dcolumn}
\usepackage{bm}

\usepackage[utf8]{inputenc}
\usepackage[T1]{fontenc}
\usepackage{mathptmx}

\begin{document}

\preprint{something}

\title[Ideal MHD coupling to the circuit]{Simulating a pulsed power-driven plasma with ideal MHD}

\author{A. Beresnyak}
\thanks{Corresponding Author}
\email{andrey.beresnyak@nrl.navy.mil}
\affiliation{Plasma Physics Division, US Naval Research Laboratory, Washington, DC 20375}
\author{A. L. Velikovich}
\affiliation{Plasma Physics Division, US Naval Research Laboratory, Washington, DC 20375}
\author{J. L. Giuliani}
\affiliation{Plasma Physics Division, US Naval Research Laboratory, Washington, DC 20375}
\author{S. L. Jackson}
\author{J. T. Engelbrecht}
\author{A. S. Richardson}
\author{A. Dasgupta}
\affiliation{Plasma Physics Division, US Naval Research Laboratory, Washington, DC 20375}

\date{\today}

\begin{abstract}
We describe a simple practical numerical method for simulating plasma driven within a vacuum chamber by a pulsed power generator. Typically, in this type of simulation, the vacuum region adjacent to the plasma is approximated as a highly resistive, light fluid; this involves computationally expensive solvers describing the diffusion of the magnetic field through this fluid. Instead, we provide a recipe for coupling pulsed power generators to the MHD domain by approximating the perfectly insulating vacuum as a light, perfectly conducting, inviscid MHD fluid and discuss the applicability of this counter-intuitive technique. This, much more affordable ideal MHD representation, is particularly useful in situations where a plasma exhibits interesting three-dimensional phenomena, either due to the design of the experiment or due to developing instabilities. We verified that this coupling recipe works by modeling an exactly solvable flux compression generator as well as a self-similar Noh-like solution and demonstrated convergence to the theoretical solution. We also showed examples of simulating complex three-dimensional pulsed power devices with this technique. We release our code implementation to the public\cite{verif_coupling}.
\end{abstract}

\maketitle

\section{Introduction}
Pulsed power plasma devices such as Z-pinches, wire arrays and dense plasma foci (DPF) \cite{Ryutov2000,Haines2011,Liberman2012} use a vacuum transmission line to transfer electromagnetic power from the capacitor bank to the conductive load. Once the electromagnetic wave reaches the load and starts pushing it, the electromagnetic field in the vacuum settles into the quasistationary state with $E\ll cB$, 
which evolves on the hydrodynamic timescale, 10–1000 ns, much longer than the electromagnetic timescale.

In this situation the displacement current
density, $\partial {\bf E}/\partial t$, can be neglected compared to ${\bf J}/\epsilon_0$ and the system can be subdivided into
two domains. The first domain, the generator itself, can be described by a model of the equivalent electric circuit\footnote{Here we assumed that various loss phenomena due to evaporation from electrode surfaces or imperfect vacuum can be incorporated into a lumped circuit model see, e.g., ref.\cite{Jennings2010}.} involving just second-order ODEs. The second domain, the conductive material in the plasma device, can be modeled using magnetohydrodynamics (MHD)---a single fluid prescription neglecting
displacement current and having material moving under pressure and Lorentz forces. The main question, therefore, is how to couple the generator circuit model to the plasma domain.

\begin{figure}[!t]
\includegraphics[width=0.6\columnwidth]{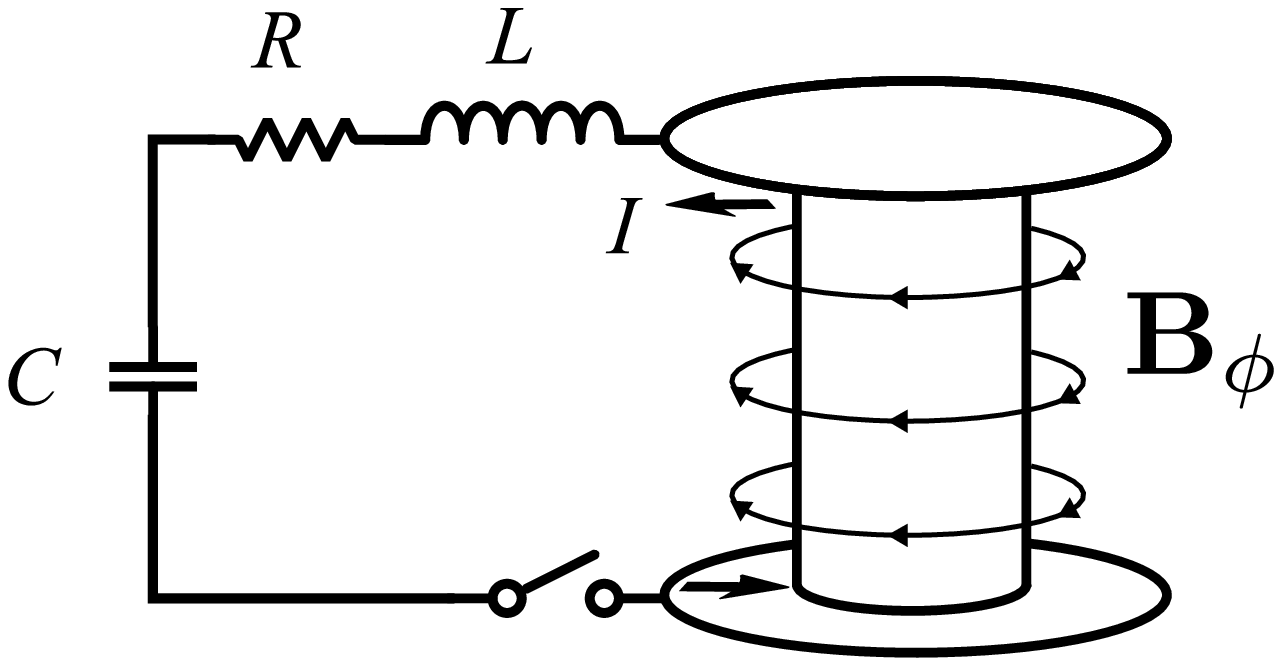}
\hfill
\includegraphics[width=0.3\columnwidth]{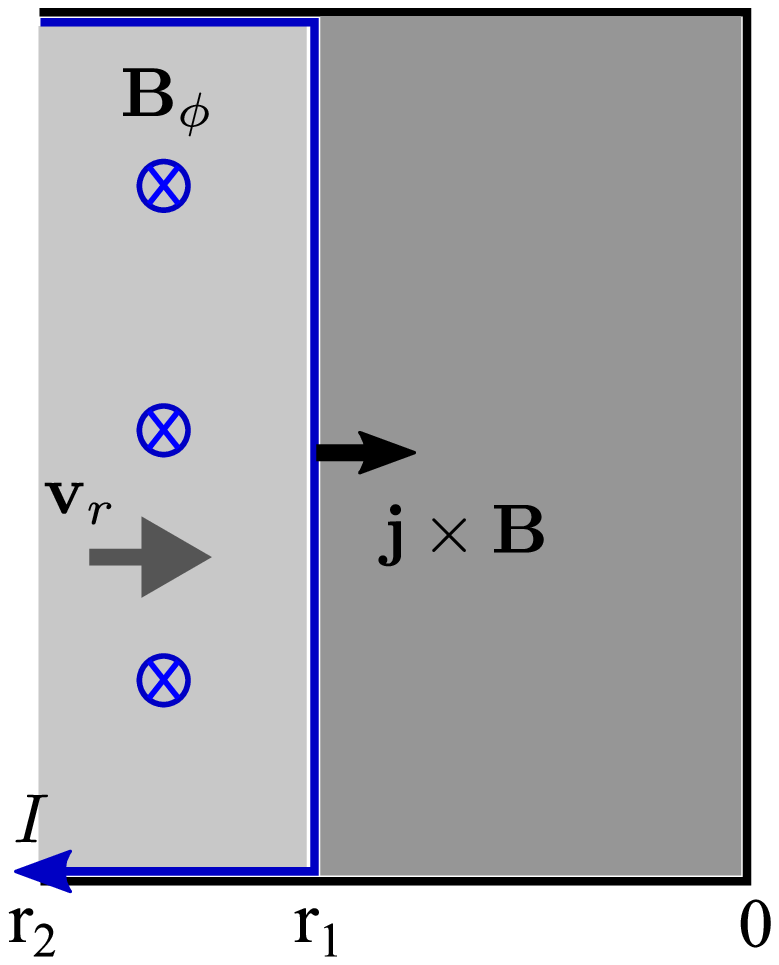}
\caption{Left: Schematics of a Z-pinch. Here, the pulsed power generator is represented by its equivalent RLC circuit. The circuit feeds current $I$ into the device, generating magnetic field $\mathrm{B}_\phi$
which compresses the load to the axis. Right: Representation of MHD domain of the Z-pinch in a simulation. The horizontal axis maps to the radial coordinate. ``Vacuum'' region is depicted light gray and cylindrical plasma load is gray. The current $I$ is flowing on the electrodes, as well as the interface between ``vacuum'' and plasma. The initial plasma velocity is, typically, set to zero, while the initial ``vacuum'' velocity is set to satisfy the induction equation as explained in \S~\ref{initial}.}
\label{z_pinch}
\end{figure}

Typically this type of plasma load is arranged as a cylinder with electrodes at either end of the plasma, see Fig.~\ref{z_pinch}. Over the course of the electrical pulse, the plasma is compressed or ``pinched'' onto the axis, leaving vacuum or low-density plasma behind at large radius.
This vacuum/low-density region cannot be properly described by MHD, with the displacement current, charge separation, and the Hall effect playing significant roles. 
These extended MHD effects are just now being investigated for their effects in Z-pinch dynamics. Within the MHD approximation, the details of the evolution of the dense plasma region can be sensitive to the treatment of the low density region, since it transmits the Poynting flux from the boundary to the dense plasma, see more on this in \S~\ref{discussion}.

Several approaches exist to a) emulate power flow from a pulsed power generator, b) model the ``vacuum'' region so that material forces are properly applied to the dense plasma, and c) 
provide feedback to the generator circuit model to use in calculating its own time-dependent circuit parameters and current. A large class of such modeling efforts involves moving mesh simulations, in which the plasma and ``vacuum'' region can be explicitly separated with a special interface. This interface is tracked in time and the condition of the power flow is applied directly to the interface.
An early example of this approach was used in modeling non-LTE ionization kinetics with a 1D Lagrangian MHD code\cite{Davis1995}. Similar scheme is used in 2D MACH2 code\cite{Peterkin1998,Peterson2004}.
 However, this approach is not without its disadvantages, especially in multi-dimensional simulations. First is that if the geometry of solid conductors in the plasma device, is complex, it may be difficult to invent a mesh that would move smoothly through such an environment. The second disadvantage is that interface tracking artificially suppresses mixing, so that when in reality the interface may become distorted and complex and leave behind a part of the plasma, such an interface will artificially suppress these effects. Among the effects due to plasma left behind are ``restrikes'' which result in a sudden decrease in inductance by introducing a return-current path in parallel with the load. The third disadvantage is that moving mesh simulations are harder to set up and are computationally more expensive than simulations based on structured Eulerian grids.

In Eulerian simulations the standard approach is to approximate ``vacuum'' regions as highly resistive low-density material\cite{Chittenden2001,Chittenden2004}, so that magnetic field quickly diffuses through the ``vacuum''. This reduces current density ${\bf j}$ and makes sure that the Lorentz force ${\bf j\times B}$ acting on this artificial vacuum is small. The approach requires fine-tuning because, if the density of the ``vacuum'' region is too high, it may, over time, obtain sizable momentum, which will upset the force balance. If the ``vacuum'' density is too low, even small ${\bf j\times B}$ forces can generate high velocities which may result in shocks and numerical heating. Highly resistive fluids are numerically expensive to simulate --- this requires either implicit solvers which are difficult to parallelize or very small time steps.
In the resistive vacuum approach, when the lumped circuit model drives MHD vacuum with both voltage and current, a special care should be taken to match the impedance of the last element of the lumped circuit to the impedance of the ``vacuum'' MHD region\cite{Jennings2010}. Another trick used in calculations of MHD fluid with low density floor is to bound Alfv\'en speed with displacement current term\cite{boris1970}. This approach, commonly utilized in magnetospheric simulations and recently implemented for ALE codes\cite{mcgregor2019}, however, requires modifications of MHD solver\cite{matsumoto2019} and can not be used with standard MHD codes. Another option to better describe vacuum in pulsed-power simulations is using extended MHD\cite{Martin2010,Seyler2011,Angus2019,Angus2020} or hybrid fluid-kinetic approaches\cite{Sefkow2019}.

\section{Rationale of Ideal MHD approach}
In this paper, we will argue that the ``vacuum'' does not have to be highly resistive. It is sufficient that we start with a specific initial state that ensures that current density ${\bf j}$ is zero in the ``vacuum'' region and that the fast magnetosonic mode in the ``vacuum'' does not couple efficiently with the motions of the dense plasma --- this typically requires that the density of ``vacuum'' is small and its magnetic field is sizable. Subsequent evolution will keep ${\bf j}$ close to zero in the ``vacuum'' region, satisfying the induction equation by maintaining the ideal Ohm's law. In practice this will require either estimating
fast magnetosonic time scale in the vacuum region analytically and choosing ``vacuum'' density based on that or adjusting ``vacuum'' density in an empirical fashion. Later in the paper we provide examples of numerically converged solutions that are not affected by ``vacuum'' density.
The physical reason
that our method works is that we are not actually interested in the physics of the artificial vacuum: the only requirement is that the information about the changing current is transferred sufficiently quickly to the vacuum-plasma interface so that plasma dynamics are described accurately. In our case, this requires that the fast magnetosonic speed is sufficiently large that the time it takes for a magnetosonic pressure wave to propagate from the entrance of the device to the plasma is much smaller than the characteristic evolution time of the plasma $\tau$. Based on that it is now straightforward to compare our approach to the resistive MHD vacuum. In the latter case, it is necessary to require that the magnetic field diffuses through the vacuum region faster than the evolution of the plasma, or the magnetic diffusivity $\eta$ is larger than $L^2/\tau$, where $L$ is a size of the vacuum region. For ideal hyperbolic codes this requirement becomes
$v_A>L/\tau$. When we plug this into resistive timestep $\Delta t_{\rm res} = \Delta x^2/ \eta$ and hyperbolic timestep $\Delta t_{\rm hyp} = \Delta x/ v_A$ we get $\Delta t_{\rm res} < (\Delta x^2/L^2) \tau$
and $\Delta t_{\rm hyp} < (\Delta x/L) \tau$. So reproducing the same simulation with higher resolution will be much more costly for the resistive code. 

Our approach has limitations --- most notably it will not work when the current feeding the plasma device goes through zero. This follows directly from the discussion above. High Alfv\'en speed is required to transfer the information about the changing current sufficiently quickly to the vacuum-plasma interface.
As we show in the next sections, the solution of the induction equation requires that the gradient of velocity in the vacuum region is $-(1/I)dI/dt$. In conventional pulsed power setting this is hardly a limitation because by the time current goes through zero the plasma dynamics becomes unstable, which makes detailed accurate simulations meaningless. 

Our approach also has a number of advantages that are particularly attractive when simulating plasma devices with complex three-dimensional geometries. One is its simplicity --- our approach is uniform. ``Vacuum'' is not treated in any special way\footnote{Our current implementation with Athena code, that is relesed to the public\cite{verif_coupling} does not interfere with the ideal MHD code solver in any way, instead it simply sets initial and boundary conditions.}, which reduces the complexity of the simulation and possible artifacts related to electrodes. As we mentioned above, in the ideal MHD regime the simulation timestep is only limited by the wave propagation speed --- typically the Alfv\'en speed\footnote{We set up ``vacuum'' fluid as cold and subsequent numerical heating does not increase the sound speed to exceed $v_A$} $v_A$. This is much less restrictive than the time step limit from the explicit resistive solver. We also argue that a resistive ``vacuum'' requires the choice of resistivity as well as density floor, which adds to its complexity. In our new approach we just have to make sure that the fluid density of the vacuum region is small enough so that in turn the $\tau$ above is small enough, and this normally also ensures that the ``vacuum'' kinetic energy does not affect the overall energetics. Our main point is that general-purpose MHD code without any special modifications regarding vacuum physics or the physics of the outside drive circuit can be used efficiently. Note, that in our case, just as in the case of resistive vacuum the vacuum density evolves in time. 
Below we will demonstrate, in two verification examples, that such an approach can simulate conservative systems rather precisely. We will also present 
a) a validation example in which we modeled the collapse of a gas puff Z-pinch,
b) a complete 3-D numerical experiment of the run-in (or sweep-up) phase of a real DPF in a complex 3-D geometry.

\section{Code description}
We used Athena \cite{Stone2008}, which is a finite volume MHD code based on Riemann solvers, supporting Cartesian, cylindrical and spherical coordinates. For the purpose of this paper, we used the HLLD flux\cite{Miyoshi2005} with PLM reconstruction in space and van Leer limiter\cite{VanLeer1974}, 2nd order time-stepping, and Cartesian and cylindrical coordinates. Athena allows for the introduction of non-ideal terms such as viscosity and resistivity, which were not used here. The Athena code is available on GitHub \cite{athena,athenapp_url}. Athena uses code units that are similar to SI, except that $\mu_0$ is equal to unity. Normally the parameters of the equivalent circuit of the pulsed power generator and its initial voltage are given in SI units, which is what our circuit solver used. 
Values of current in SI units were translated to Athena units, and the values of voltage, calculated by Athena, were translated to SI units to feed back to the circuit solver.  Athena++ is capable of adaptive grid refinement, which we do not use for the purposes of this paper, however, our method may be easily implemented on adaptive grids as well.

\section{Initial and boundary conditions, time integration}
\label{initial}

The initial condition for the plasma domain usually implies zero velocity, or velocity close to zero. In this situation, the change in the magnetic field in the ``vacuum'' region is entirely due to the rate of change in the circuit. We also would want to satisfy the induction equation,

\begin{equation}
    \frac{\partial \bf B}{\partial t} = - {\bf \nabla \times E}, \label{induct}
\end{equation}
where the change of $B$ will be determined by the $dI/dt$ in the circuit, while the electric field will assumed to be $-{\bf v \times B}$, the ideal MHD field. In general, this will require solving the Laplace equation for magnetic potentials or equations ${\bf \nabla \cdot B}=0$, ${\bf \nabla \times B}=0$, ${\bf \nabla \cdot E}=0$, ${\bf \nabla \times E}=0$ with boundary conditions specifying ${\bf \nabla \times B}$ from the current-carrying surfaces and the boundary conditions specifying ${\bf \nabla \times E}$ at the entrance of the device. However, in many cases, the solution is trivial, as it is for the verification problems that we investigate further in this paper.
So the ``vacuum'' region is set according to three conditions as depicted in Fig.~\ref{z_pinch}: a) the density is set to a small value that is still affordable in terms of the Alfv\'enic time step constraint, note that $\Delta t \sim \sqrt{\rho}$; b) the magnetic field is set to satisfy ${\bf \nabla \times B}=0$ in the ``vacuum'' region with appropriate boundary conditions for the current flowing on the surface of the device as well as on the surface of the plasma. For example, in the cylindrically symmetric case
such a field is given by $B_\phi=\mu_0I/2\pi r$; c) The velocity is set to satisfy Eq.~\ref{induct} (see, e.g., Eq.~\ref{vx_cart}).

The boundary condition imposed by the generator on the MHD domain was set as a free-flow with an important modification. The velocity in the ghost zones outside of simulation domain was set as a linear function from the first zone within the simulation domain with the velocity gradient determined by $dI/dt$ of the circuit in a way similar to the procedure for the initial condition. Note that this involves first order interpolation
in boundary velocity. The particular solution that we use for the boundary condition in the Cartesian geometry is $dv/dx=-(1/I)dI/dt$ and in cylindrical geometry it is $dv/dr=v/r-(1/I)dI/dt$.
The density in the ghost zones was set to ``vacuum'' density. The voltage on the device, which is needed to solve the circuit equation, is determined from the $-{\bf v \times B}$ electric field in the first zone in the MHD domain.

The initial condition for the circuit involves the initial current as well as the initial charge on the capacitors, but the MHD portion of the problem will also require an initial $dI/dt$. This initial $dI/dt$ is obtained by equating the initial voltage at the output of the circuit to $L_{\rm vac} dI/dt$ and solving for $dI/dt$. $L_{\rm vac}$ is the initial static inductance of the ``vacuum'' region. The inductance is static because the initial plasma velocity is zero and the walls of the device are also at rest.
Subsequently, the circuit equation is solved using voltage calculated from the MHD dynamics.
We would like to emphasize that the voltage on the device is represented as the static vacuum inductance only on the first time step, e.g., it may be zero if the initial ``vacuum'' volume is zero. However, in the subsequent self-consistent calculation, the voltage on the device can no longer be described as inductive due to the non-trivial evolution of the plasma.

The MHD problem and the circuit equation (the equation for $I$) are integrated simultaneously, with the time step determined by the MHD code.
A first order Euler solver is generally good enough for the circuit, since the MHD time step is usually constrained to be small compared to the evolution time of the circuit. More quantitatively, typically for pulsed power experiments, MHD timescale of the plasma are shorter or equal to the circuit timescale, while the MHD time step is determined, roughly, as MHD timescale divided by the number of grid points in one direction.

\begin{figure}[!t]
\centering
\includegraphics[width=0.8\columnwidth]{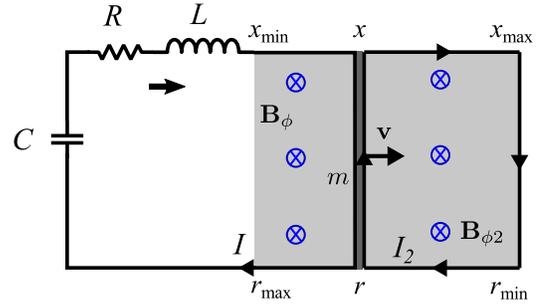}
\caption{Verification problem, MHD domain is shown in gray, also shown: a) an equivalent RLC circuit of the generator, b) two MHD ``vacuum'' domains, shown in light gray, with currents $I$ and $I_2$ circulating on the surface and c) the mass $m$, shown in dark gray, residing at a coordinate $x$, which is subjected to magnetic pressure from both vacuum domains. In cylindrical case the liner is pushed from large radius $r_{\rm max}$ never reaching smaller radius $r_{\rm min}$, where conductive boundary is situated. Likewise, in Cartesian case the liner slides back and forth between $x_{\rm min}$ and $x_{\rm max}$.}
\label{flux_compr}
\end{figure}

\begin{figure}[!t]
\centering
\includegraphics[width=0.99\columnwidth]{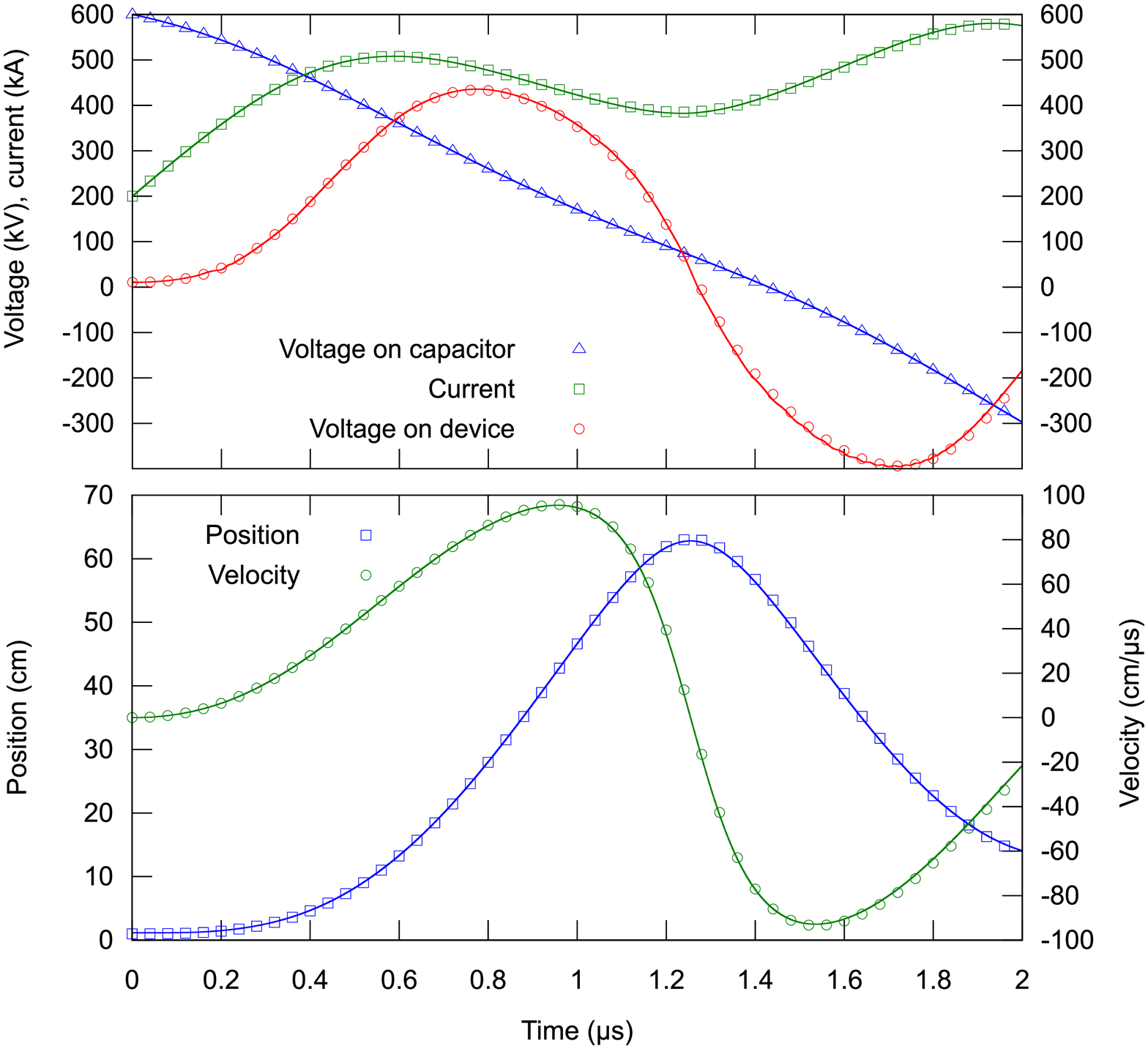}
\caption{Comparison of theoretical solution (symbols) with numerical ideal-MHD solution (solid lines) for the Cartesian geometry verification problem. On the top panel, we show circuit variables, on the bottom --- position and velocity of the mass.}
\label{verif_z}
\end{figure}

\section{Flux compression verification problems}
\label{ver_ode}
The utility of the method for modeling the coupling between a pulsed power generator and a plasma device has been verified by applying it to a specific, easily integrable system and comparing its result to the exact analytical solution\footnote{We release these simple implementations of our coupling method to the public\cite{verif_coupling}.}.
 For this purpose, we selected a convenient problem 
in which all the mass of the plasma is concentrated into a thin liner, the dynamics of this mass are one-dimensional, and finally, the dynamics should not produce strong shock waves. The last condition ensures that if we set initial specific entropy to zero, it will stay close to zero during the evolution. We keep in mind that a realistic computer simulation will produce numerical heating, but we will argue that, in the absence of shocks, the influence of numerical heating and numerical diffusion will be decreased by increasing the resolution of the numerical experiment. The particular problem we choose is ``flux compression'' --- when the magnetic field produced by the current from the generator provides pressure on the plasma from one side, while the magnetic field, circulating within the device, will provide counter-pressure. As the liner is initialized with the same thermal pressure, subsequent evolution does not create shocks.
Such a conservative system is simply a nonlinear coupled oscillator. It can be
described with ordinary differential equations and solved very precisely.
To model the pulsed-power generator, we use a serial RLC circuit. This problem is visualized in Fig.~\ref{flux_compr}.

Instead of choosing a specific coordinate system, let us designate the inductance of the first vacuum region as $L_1(x)$ and the inductance of the second vacuum region as $L_2(x)$. We denote initial position as $x_0$ (or $r_0$) and initial currents as $I_0$ and $I_{20}$. In this case the current and velocity of the mass $m$
can be evolved via the two equations:

\begin{equation}
    I\frac{\partial L_1(x)}{\partial x}v+(L+L_1(x))\frac{dI}{dt}=\frac{q}{C}-IR, \label{current_evol}
\end{equation}
\begin{equation}
    m\frac{dv}{dt}=\frac 12 \frac{\partial L_1(x)}{\partial x}(I^2-I_2^2), \label{v_evol}
\end{equation}

where $I_2$ is determined from the flux conservation constraint: $I_2=I_{20}L_2(x_0)/L_2(x)$. Adding the definitions $I=-dq/dt$ and $v=dx/dt$ closes the system with four equations and four dynamic variables: coordinate of the liner $x$, its velocity $v$, charge on capacitor $q$, and current $I$. Note that our numerical scheme from Section~\ref{initial} requires that $v=0$ initially.

The general problem described above can be specified in different geometries. 
We have chosen to solve the problem for two cases: 1) a planar, translating liner in a cartesian, $x-z$ coordinate system, and 2) an annular, imploding liner in a cylindrical, $r-z$ geometry. For the cylindrical geometry, the cartesian coordinate, $x$, that points in the direction of the plasma's motion is replaced in Eq.~(\ref{current_evol}) and (\ref{v_evol}) by the radial coordinate, $r$.

In the Cartesian case:
\begin{equation}
    L_1(x)= \mu_0 \frac{l_z}{l_y} (x_{\rm max}-x), \ 
    L_2(x)= \mu_0 \frac{l_z}{l_y} (x-x_{\rm min}).  
\end{equation}

In the cylindrical case:
\begin{equation}
    L_1(r)= \frac{\mu_0}{2\pi} l_z \ln \frac{r_{\rm max}}{r}, \ 
    L_2(r)= \frac{\mu_0}{2\pi} l_z \ln \frac{r}{r_{\rm min}}.  
\end{equation}

\begin{table}[!t]
\caption{Parameters of the simulations}
\begin{tabular*}{\columnwidth}{l @{\extracolsep{\fill}} c c}
    \hline\hline
Simulation & Cartesian & r-z \\
\hline
y-z dimensions & $l_z/l_y=1$  &  $l_z=60$ cm \\
$x_{\rm min}$ & 0 cm & 8 cm \\
$x_0$         & 1 cm &  18 cm \\
$x_{\rm max}$ & 77.9 cm &  20 cm \\
m & 80 $\mu$g & 80 $\mu$g \\
$\rho_0$ & $10^{-4}\, \mu\rm g/cm^3$ & $10^{-7}\, \mu\rm g/cm^3$ \\
$N_x$ & 400 & 800 \\
   \hline
\end{tabular*}
  \label{experiments}
\end{table}

In Cartesian geometry the solution of the Eq.~(\ref{induct}) at $t=0$ will be expressed as 
\begin{equation}
    v_x=\frac 1 I \frac{dI}{dt}(x_0-x).
    \label{vx_cart}
\end{equation}
This is used to set the initial condition in the left vacuum region. 
Since we are assuming that the initial plasma mass velocity is zero, then
$dI_2/dt=0$ and $v_x$ for the vacuum region on the right is identically zero at $t=0$. 
Then we proceed to solve a coupled MHD-circuit problem as described in Section~\ref{initial}.

Figure \ref{verif_z} presents a comparison of the ODE solution of the idealized system described by Eq.~(\ref{current_evol}) and (\ref{v_evol}) and the MHD solution obtained by our coupling method from Section~\ref{initial}. The parameters of the problem are described in Table~1. Other parameters include the parameters of the circuit, which is the same in both verification problems: $C=1\,\mu$F, $L=700$ nH, $R=0\,\,\Omega$, initial voltage on capacitor --- $V_0=600$ kV, initial current --- $I_0=I_{20}=200$ kA.
\begin{figure}[!t]
\centering
\includegraphics[width=0.99\columnwidth]{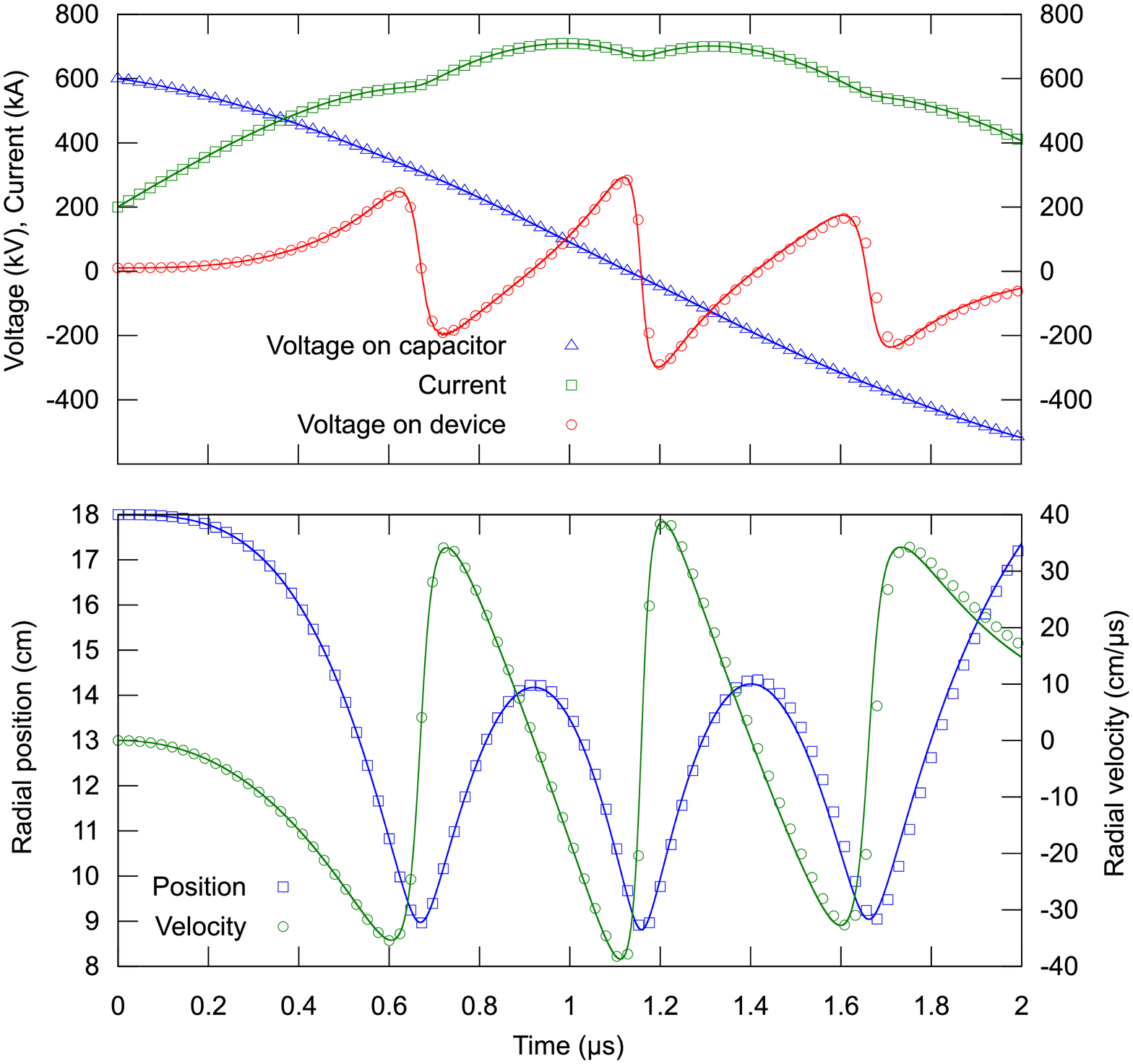}
\caption{Same as Fig.~\ref{verif_z}, but for cylindrical verification problem.}
\label{verif_r}
\end{figure}

As we mentioned above, provided that $R=0\,\,\Omega$, this system is conservative and the nature of the solution is oscillatory. In the beginning, the growth of $I$ pushed the mass to compress inside magnetic field and after some time it rebounds back under the pressure of the compressed flux inside the right vacuum region.


In the MHD simulation the material of the liner is subject to numerical diffusion and is spread over many cells. In this case the position and velocity of the liner that corresponds to variables $x$ and $v$ from Section~\ref{ver_ode} have to be defined in the MHD domain. We defined them
as the center of mass and the velocity of the center of mass of the whole volume.

Fig.~\ref{convergence}, top panel, presents a numerical convergence of the theoretical values from the ODE solver to numerical values. This convergence is somewhat worse than first order, which we think is because our plasma density is represented as a point mass, a delta function on a grid. Likewise, current density is also represented as a delta function in the theoretical picture on Fig.~\ref{flux_compr}. The effects of mass diffusion as well as numerical heating are indeed prominently seen in numerical solution.

In r-z geometry the solution of Eq.~(\ref{induct}) at $t=0$ will be expressed as 
\begin{equation}
    v_r=-\frac 1 I \frac{dI}{dt}r \ln \frac{r}{r_0}.
\end{equation}
Similar to the Cartesian case, we set the velocity to this initial value in the vacuum region outside the liner, while $v_r=0$ inside initially.
We proceed to solve a coupled MHD-circuit problem 
in a manner similar to the above section.
\begin{figure}[!t]
\includegraphics[width=0.9\columnwidth]{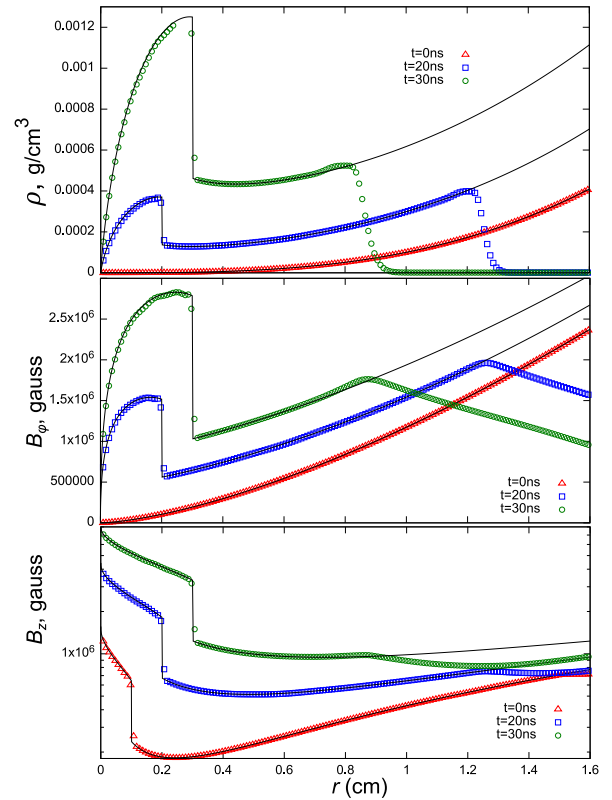}
\caption{Mag Noh numerical solution (points) vs exact self-similar solution (black lines).}
\label{magnoh}
\end{figure}

We present verification results in Figure~\ref{verif_r}. Note that the collapse and rebound motion
of the mass is very fast and produces relatively short dips in the total current. It is due to this very fast motion that the r-z problem is more challenging and typically requires lower ``vacuum'' density.

\section{Mag Noh verification problem}
Our flux compression verification tests were fairly idealized compared to actual Z-pinches as they assume that plasma is, essentially, a point mass. The convergence to high accuracy was also slow because our solution, formally, had a mass distribution in a form of a delta function in $x$ or $r$, which was subject to fast numerical diffusion. Recently we developed a suite of exact self-similar solutions of MHD equations with a fast MHD shock, resembling plasma driven by magnetic field in Z-pinches\cite{velikovich2012,beresnyak2021}. We released the code to produce such solutions to the public\cite{magnoh_code} and already used these solutions to test stability and accuracy of various codes, such as Athena, Mach2 and Flash. For the purpose of this paper we assume that the plasma, representing the self-similar Mag Noh solution in cylindrical geometry, is being pushed in and compressed by low density artificial vacuum similar to what we described above. However, in order to maintain the necessary ${\bf j\times B}$ force we will have to adjust the feed current so that it corresponds to the magnetic field in the Mag Noh solution. In this case the current source can not be represented as a simple LRC circuit, however the coupling method that we use on the boundary is essentially the same as in previous problems. Below we describe the initial
condition as well as a boundary condition for this problem.
\begin{figure}[!t]
\includegraphics[width=0.9\columnwidth]{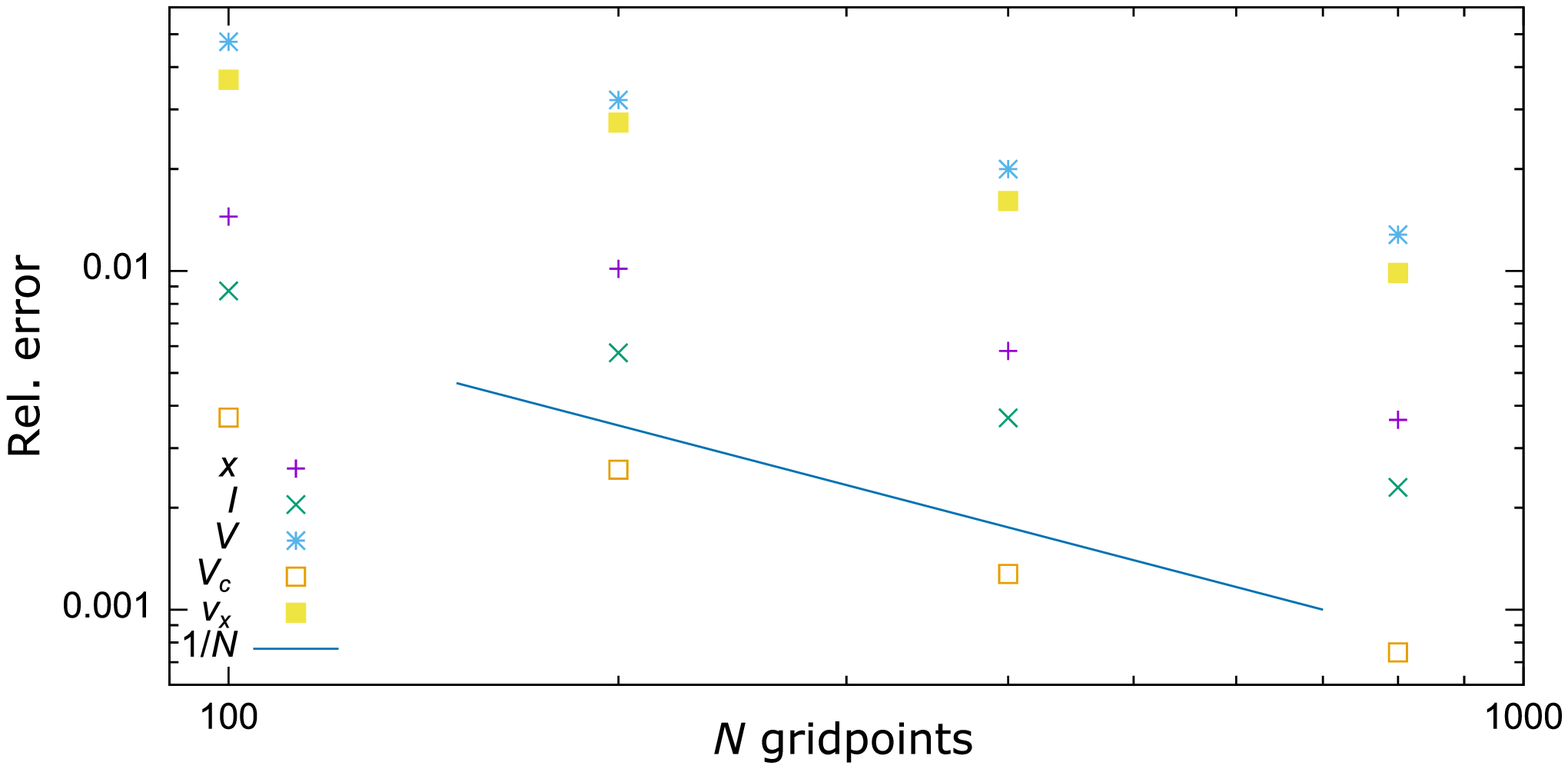}
\includegraphics[width=0.9\columnwidth]{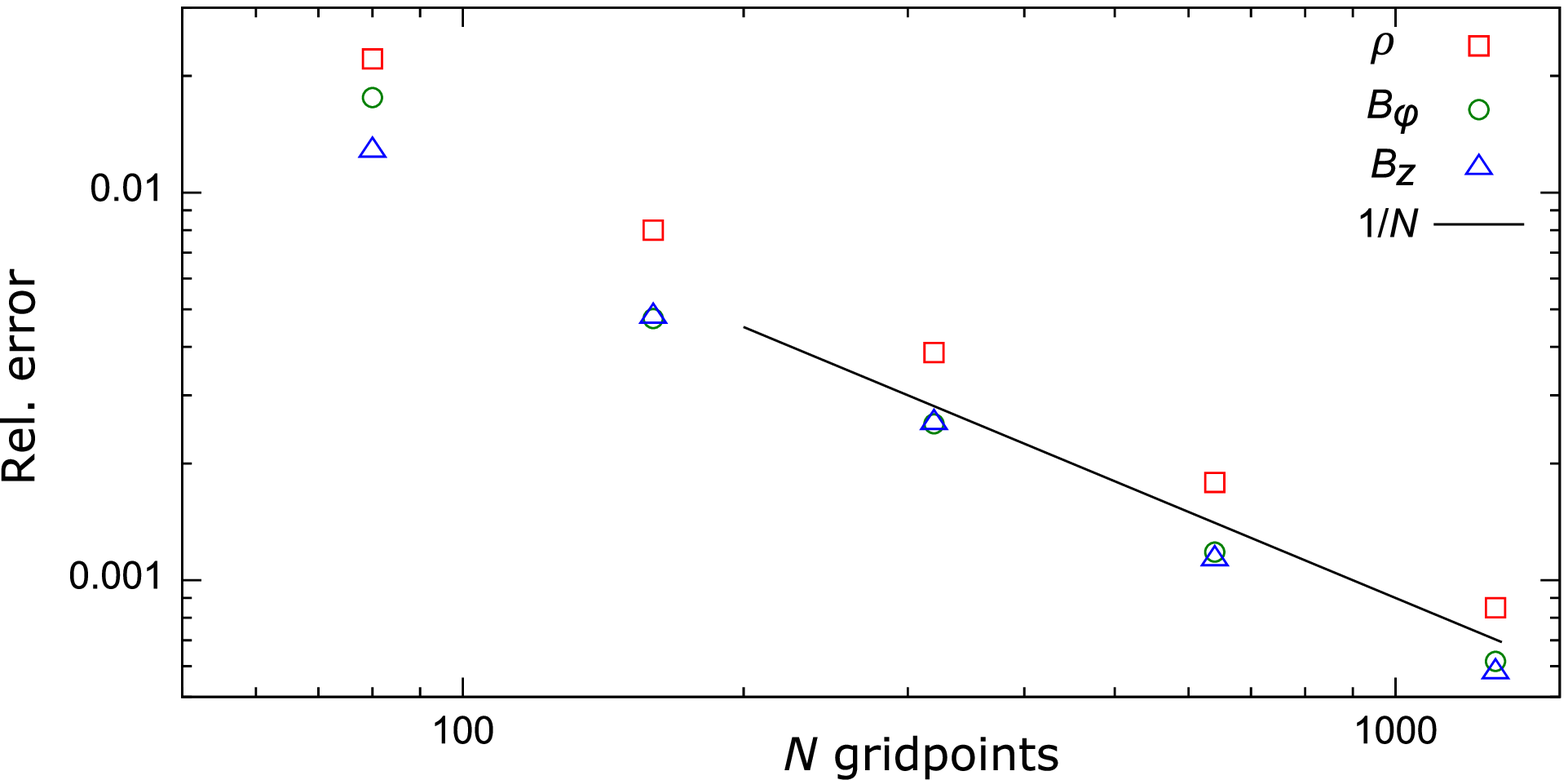}
\caption{Top: Numerical convergence of the flux compression problem. We plot root-mean-square (RMS) deviation for each quantity in the whole time interval divided by the absolute maximum of that quantity. Bottom: Numerical convergence of the Mag Noh solution at $t=30$ns. Similar relative RMS error but over the interval of radii.}
\label{convergence}
\end{figure}

We initialize plasma in the cylindrical r-z numerical domain with zero initial velocity, zero pressure, density and magnetic field as power-laws like $\rho=10^{-4} (r/1{\rm cm})^3 {\rm g/cm}^3$, $B_\phi=1.17\times 10^6 (r/1{\rm cm})^{1.5} {\rm gauss}$,
$B_z=3.52\times 10^5 (r/1{\rm cm})^{1.5} {\rm gauss}$. This initial condition is expected to produce a shock with a constant
shock speed of $10^{7} {\rm cm/s}$ due to acceleration of plasma by magnetic tension. Initial volume of the artificial vacuum is zero. On the boundary we apply an inflow condition for velocity similar to above, with a ``vacuum'' density of $3\times 10^{-7} {\rm g/cm}^3$ and we set the boundary condition for the magnetic field as described below. The location of the ``edge of the plasma'', $r_e(t)$ is calculated by integration of velocity $v_{ss}(r_e,t)$ from the self-similar solution, $r_e(t)=\int v_{ss}(r_e(t),t) dt$. The boundary values for magnetic field are set to reproduce ideal vacuum solution $B_\phi \sim 1/r, \, B_z \sim {\rm const}$. In our case the boundary values are matched to the values on the edge of the plasma $r_e$: $B_\phi(r_{\rm max})=B_{\phi ss}(r_e,t) r_e/r_{\rm max}$ and $B_z(r_{\rm max})=B_{z ss}(r_e,t)$.
Here we designated the self-similar solution with subscript ``ss'': $v_{ss}(r,t),\, B_{\phi ss}(r,t),\, B_{z ss}(r,t)$, these values are taken from the table. 
The self-similarity of these functions is expressed in the following way -- for any two $r_1,t_1$ and $r_2,t_2$ satisfying $r_1/t_1=r_2/t_2$ we have $v_{ss}(r_1,t_1)=v_{ss}(r_2,t_2)$, 
$ B_{\{\phi,z\} ss}(r_1,t_1)= (t_1/t_2)^{1.5} B_{\{\phi,z\} ss}(r_2,t_2)$, 
$ \rho_{ss}(r_1,t_1)= (t_1/t_2)^3 \rho_{ss}(r_2,t_2)$. For more detail see \cite{velikovich2012,beresnyak2021}.
Our solution, therefore, is a self-similar solution stitched to the vacuum solution.

Note that we put our method to additional test by introducing the $B_z$ field. Solutions with zero $B_z$ are also available\cite{magnoh_code} 
and can be tested. We can now evaluate whether the numerical plasma values are close to the theoretical self-similar solution values and whether the artificial vacuum values are close to the vacuum solution. The results are presented on Fig.~\ref{magnoh}. We see that the high-density region occupied by the plasma exhibit
a solution which is close to a self-similar solution. At the same time, the physical quantities in the artificial vacuum region deviate from the vacuum $B_\phi \sim 1/r, \, B_z \sim {\rm const}$, which is not surprising, since our artificial vacuum has a finite density and it takes a finite Maxwell stress to accelerate artificial vacuum to match the velocity at the edge of the plasma. Our method still robustly produces accurate plasma solution because it transmits the accurate amount of Pointing flux through the plasma-vacuum boundary. Figure~\ref{convergence}, bottom panel, presents a numerical convergence of the plasma solution between 0 and 0.6cm and excluding the shock to the theoretical self-similar solution. This convergence is slightly better than first order.

\begin{figure}[!t]
\centering
\includegraphics[width=0.99\columnwidth]{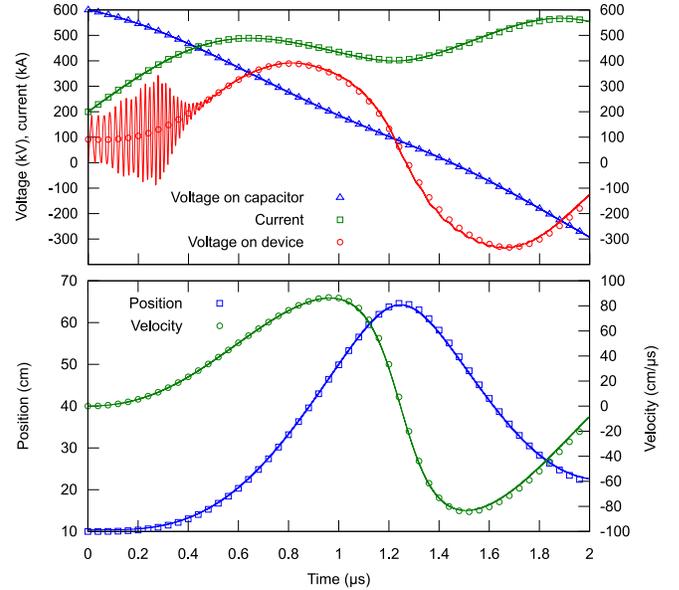}
\caption{Cartesian verification problem with $\bf E=0$ set initially, which is inconstistent with vacuum electric field from Eq.~\ref{induct}. 
The voltage on the device exhibits rapid oscillation due to a fast wave traveling between entrance of the device and the mass.}
\label{verif_z_bad}
\end{figure}
\section{Caveats of the coupling method}
\label{caveats}
From the above verification problems, we have established that our coupling method will correctly reproduce the theoretical solution if used properly.
One of the ways this coupling method can develop artifacts is when the initial electric field in the MHD domain is not exactly matched to the theoretical vacuum electric field. For example, if the initial current is zero, so is the initial magnetic field, in which case ideal MHD electric field $-{\bf v \times B}$ must also be zero. This can not be matched to the solution of the Maxwell's equations in vacuum, which will give non-zero electric field if $dI/dt$ is not zero.
\section{Comparison with Hawk DPF experiment}
\begin{figure}
\centering
\includegraphics[width=0.99\columnwidth]{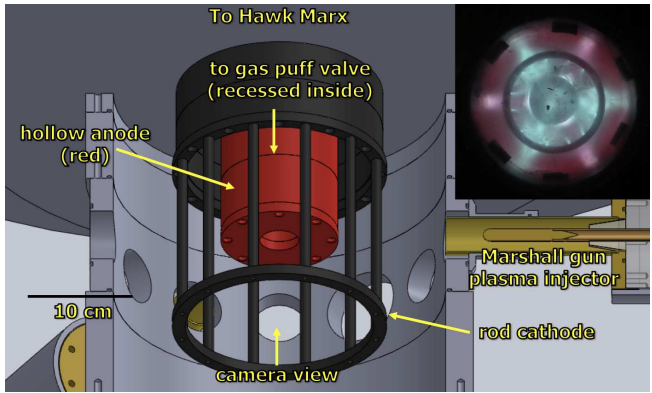}
\caption{Schematic of the Hawk DPF experiment (side view). The upper right inset shows the end-on visible-light image of plasma injected from three plasma guns into the device.\label{hawk_dpf}}
\end{figure}

If such a mismatch occurs at the beginning of the simulation the ``vacuum'' region will develop a fast shock, that will propagate within ``vacuum'' region, reflecting from the plasma and the entrance of the device many times. This will produce periodic oscillation on the velocity of the ``vacuum'', as well as the voltage on the device. Note that, according to our coupling definition, the voltage on the device in the ideal MHD case is entirely determined by the inflow velocity of the ``vacuum''. If these fictitious voltage oscillations are large enough to interfere with the circuit solution, then the overall solution will not be usable. Otherwise, the voltage oscillation will only marginally affect the dynamics of the plasma. This is often the case when simulating in a realistic setting. Furthermore, the realistic device setup may include low-density plasma left behind the magnetic piston, for example, a neutral gas left behind may be partially ionized by the radiation of the main shock. In this case, the fast wave propagating in low-density plasma would be real. In real devices, short-period voltage oscillations are often observed, but their origin is not entirely clear. Figure~\ref{verif_z_bad} shows a simulation 
with the initial state of the vacuum set incorrectly --- we set $v_x=0$, i.e., $E=0$. This generated a fast wave, which decayed over the course of multiple reflections on the receding liner.
These oscillations reappear at $t=1.3\,\mu$s with much smaller 
amplitude, after being amplified when the liner rebounds and compresses the outer portion of the ``vacuum''. Since the oscillatory fast wave carries relatively
little pressure, its signature is only noticable in the voltage on the device,
while material dynamics, reflected in position and velocity of the liner, is relatively unaffected, see Figure~\ref{verif_z_bad}.
 
To summarize, our approach requires that we set the velocity of the fluid in the vacuum region to a value that produces an electric field that agrees with the true vacuum solution. Alternatively, we can minimize the initial volume of the vacuum, so that the fictitious waves caused by mismatch bounce around very quickly and dissipate. Setting the correct initial value may require solving the Laplace equation for the vacuum region --- which may be supplied by a separate code module. We also should be careful not to excite an MHD fast wave inside of the ``vacuum'' region due to fast motion of the plasma, especially if the plasma motion is a compressive motion.  
This can be achieved by setting the vacuum density sufficiently low so that the Alfv\'enic timescale of the vacuum is much shorter than the evolution time of the plasma.
\begin{figure}[!t]
\centering
\includegraphics[width=0.99\columnwidth]{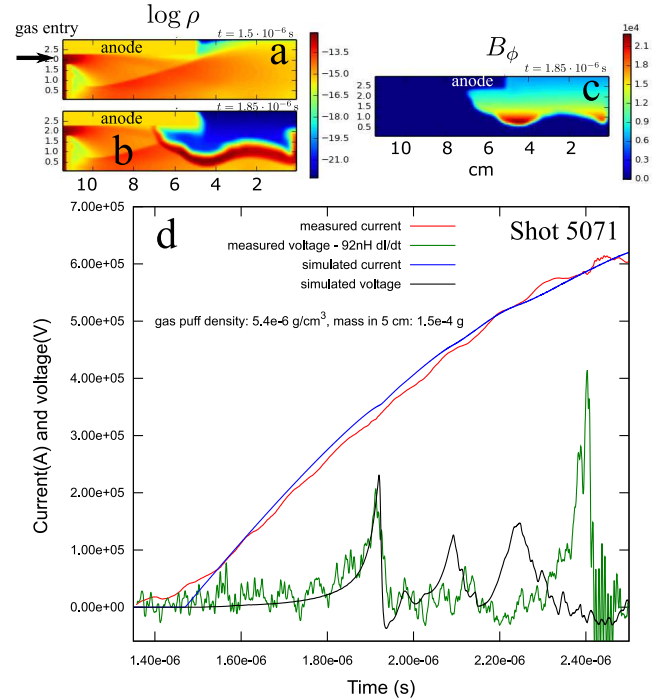}
\caption{Validation of the model by comparison with Hawk DPF experiment. 
Panel (a) shows the initial distribution of simulated gas puff density, showing a complex density structure resulting from the gas outflow. Panel (b) and (c) shows simulated density and magnetic field shortly prior to pinch. The initial distribution is not uniform along $z$, and neither is the pinch collapse, affecting the voltage signal. Panel (d) shows voltage and current traces recorded in the experiment as well as simulated by the code. We subtracted $L_{m} dI/dt$ from measured voltage to account for a constant vacuum inductance $L_{m}\approx$ 92 nH between the actual measuring location in the device --- way upstream of plasma region depicted on panels (a)-(c) --- and the measuring location in the simulation.}
\label{validation}
\end{figure}

The real-world utility of the vacuum coupling recipe was verified by applying it to simulate a pulsed-power-driven plasma experiment and then comparing the time-dependent current and voltage from the simulation to the same experimentally-measured quantities. The experiments were conducted using the Naval Research Laboratory's Hawk pulsed-power generator \cite{Commisso1992} to drive an unusual arrangement of a DPF load. As shown in Fig.~\ref{hawk_dpf}, the experiments have a fairly complicated setup, including a hollow anode, cathode with rods, local plasma injection with plasma guns and neutral gas injection with a gas puff \cite{Engelbrecht2019,Klir2020,Klir2020a}.
The plasma guns are used to inject plasma into the space between the coaxial electrodes. The gas puff is used to provide gas along the rotational axis of the experiment. Typically the plasma guns and gas puff are used together to initialize the DPF before the application of the current pulse from the pulsed power generator. But either can also be used individually.
A variety of data have been collected, including current and voltage traces for comparison with the simulation. 

One can hope that the MHD description may be applicable in the first phase of the experiment, when the plasma column or Z-pinch is formed and before the disruption of the pinch. After the disruption, not only the dynamics become unstable and more sensitive to the initial conditions, but we also see plasma generating non-ideal electric fields, breaking applicability of MHD. The indirect evidence of strong fields has been obtained in the form of accelerated ions and high-energy neutrons and will be described elsewhere. In this paper,
we try to validate our MHD model with the recorded current and voltage traces from the Hawk DPF experiment.
Figure~\ref{validation} shows the results of the validation experiment between gas-puff only
Shot 5071 on Hawk and the 2-D cylindrical MHD simulation. In this experiment the gas puff density had not been measured directly.
Instead, the normalized profile of the density from the nozzle was simulated and then given an absolute value so that the timing of the first peak in voltage in the simulation and experiment coincide. Simulation of the gas puff is described briefly in a later section of this paper.
The voltage amplitude and the profile have not been fit in any way, however. 
We have chosen to simulate Shot 5071, which had no plasma from plasma guns and pinched early, before the peak of the current, for a specific reason. This was to estimate the gas puff density and provide a benchmark value for other experiments including those with plasma guns.

As we see from Figure~\ref{validation}, the simulation reproduced the amplitude and the shape of the first voltage peak reasonably well, keeping in mind the uncertainty in the voltage measurement. The comparison of the current profiles between the simulation and the experiment primarily validated the equivalent circuit parameters of the generator, since because of the high generator impedance, it was relatively unaffected by the plasma dynamics. Following the voltage after the first pinch, we see that the simulation and experiment are starting to deviate significantly. As the simulation shows a clear rebound of the pinch at $t=1.93\,\mu$s evidenced by a negative voltage, the experiment does not show such a clear signature. Secondary and tertiary collapses of the pinch at $t=2.1\,\mu$s and $t=2.25\,\mu$s in the simulation are also not observed in the experiment. Finally, at $t=2.4\,\mu$s the experiment exhibits a complete disruption of the pinch with a high voltage spike, by which time the ideal MHD limit becomes completely invalid. Apparently our axially symmetric ideal MHD numerical experiment cannot reproduce such a spike. To summarize, the dynamics of the plasma before and during the first collapse appear to be reasonably well described by ideal MHD. After the first collapse, the dynamics are not well described by the model. This is due to a high sensitivity to the initial conditions due to Rayleigh–Taylor instability at stagnation, as well as plasma effects beyond ideal MHD. 

We should note parenthetically that the Hawk DPF simulations also included optically thin radiative cooling using cooling tables from FLYCHK and PrismSpect software\cite{chung2005,macfarlane2003}. The effect of this cooling on the plasma dynamics was negligible, however, in the context of the Hawk DPF.

\section{Three-dimensional MHD experiments}
\begin{figure}[!t]
\centering
\includegraphics[width=0.99\columnwidth]{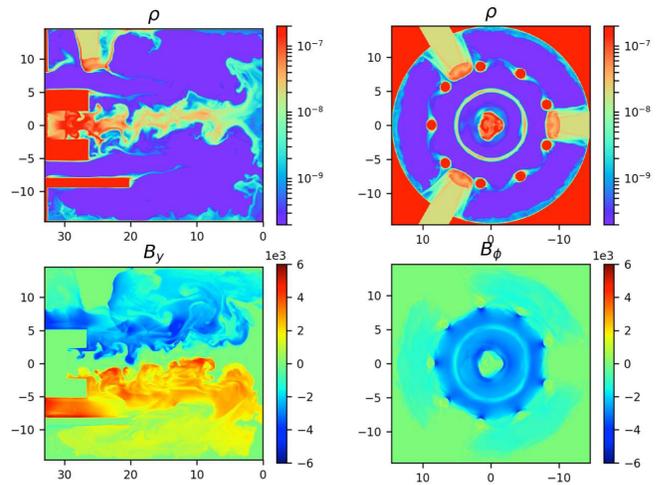}
\caption{Three-dimensional simulation of the Hawk DPF experiment, all sizes are in cm. On the left we show a two-dimensional slice along the axis, while on the right the slice is perpendicular to the axis at 2 cm from the end of the anode.}
\label{hawk_3d}
\end{figure}

From the beginning, we set as our goal to achieve a simple, verifiable way to couple an ideal MHD numerical experiment with a pulsed power generator. A big part of this goal was to ensure that the resulting combined code is sufficiently fast 
for enabling us to perform complex three-dimensional numerical simulations. These simulations will adequatly describe the complexity of experiments, such as the Hawk DPF experiment. Figure~\ref{hawk_3d} presents several slices from such a complex three-dimensional simulation. We present this simulation for illustrative purposes only, to demonstrate the capability of the code. The first phase of the numerical simulation was a pure hydrodynamic simulation of the gas puff, which injected neutral gas from the annular hole along the axis and had
longer dynamical times and much lower velocities than the subsequent pulsed power simulation. To keep the code simple we did not run this phase separately; rather it was one part of the continuous MHD evolution, but during this part the external current was set to zero.
The next stage was the injection of plasma from the plasma guns that was set at a constant velocity, which was actually measured in the experiment. The plasma from the guns collided with the gas on-axis and created a complicated density structure upon which the current from the pulsed power generator was conducted. We should notice that all gas in this numerical simulation was treated as conductive ideal gas --- that is, we did not explicitly follow the ionization of the gas from the gas puff. In 2.1 $\mu$s after the injection of plasma from the guns, the generator was fired, meaning in the code we set a non-zero initial voltage of the capacitor. Initially, the vacuum region was very small because the plasma from the plasma guns diffused and filled most of the space close to the entrance. Also, this region had a complicated shape. In this situation we decided not to solve for the initial condition of the vacuum, but set its velocity to zero, as we did in Section~\ref{caveats}. The oscillations due to mismatch of the vacuum and the entrance were insignificant. We used a small non-zero $I$ to avoid numerical complications with the very high inflow velocity associated with the boundary conditions. When the current pulse was applied, the plasma from the plasma guns was swept away from the anode and towards the axis, as is usual for DPF experiments. 
In the density plots at the top of Fig.~\ref{hawk_3d}, we see that the plasma has been swept away from the space between the anode and rod cathode and replaced by numerical vacuum. The magnetic field is depicted on the lower two plots of Fig.~\ref{hawk_3d}.

Despite the relatively high numerical resolution, $672\times 672\times 768$, due to the simplicity of the ideal MHD code, which had low communication overhead, it was possible to perform this simulation with fairly modest computing resources --- 18 hours on 128 nodes of LANL Grizzly (two Xeon E-5 per node).

\section{Discussion}
\label{discussion}
In this paper, we demonstrated that there is a numerically efficient way to simulate a plasma device coupled to a pulsed power generator. Our main point was that, despite the fact that plasma is pushed away and replaced with the magnetized vacuum in real devices, the vacuum may as well be represented by a light, ideally conductive MHD fluid in simulations. This may reduce the overhead spent on simulating the vacuum region compared to the use of non-MHD prescriptions or resistive MHD, especially for high resolution simulations. Resistive MHD is numerically expensive due to either the small explicit time step required or poorly parallelizable global implicit solvers. We verified our scheme by using ODE solvable cases, such as flux compressors in Cartesian as well as cylindrical geometries.

We noted that our method requires that the initial state has to be carefully prepared, which may be non-trivial numerically speaking. Mathematically this is, in some sense trivial, because it relies on the solution of Maxwell's equations in vacuum being fit with the ideal Ohm's law ${\bf E = -u \times B}$. 
If the initial state is not well prepared, the voltage signal may exhibit large oscillations due to a fast shock propagating in the artificial vacuum. Technically, the same thing happens in the resistive vacuum method, but such wave is quickly dissipated, that is why setting exact initial conditions for vacuum was considered not a requirement for the resistive vacuum method. And perhaps that is one of the reasons why
the resistive vacuum method had been popular. In our case the wave is still dissipated, but by numerical
diffusivity. 

Our interest in applying MHD to pulsed power devices is primarily for 
the purpose of simulating the initial stage of the plasma sweep-up and pinching, where the MHD approximation is largely applicable.
We can simulate Z-pinches, wire arrays and DPFs
in this way in order to determine the optimum plasma load configuration for coupling to the generator.
At the same time, the MHD evolution will set the stage for the subsequent pinch disruption and other non-MHD effects \cite{Anderson1958,Mather1965}. In DPFs, in particular, there is a sweep-up phase \cite{krishnan2012} which is fairly long and is too expensive to simulate with particle-in-cell codes. 
In the present approach we can still introduce resistivity to study non-ideal effects such as pinch disruption\cite{proc_dzp}, and we can simply use a density floor in the ``vacuum'' region as a threshold above which resistivity vanishes.

In this paper, 
we not only verified that our method gives accurate, convergent solutions, but also that it was successful in simulating complex three-dimensional plasma devices using a fixed grid and ideal MHD.
Future research is needed to understand the importance of non-MHD effects in the evolution prior to the pinch\cite{proc_dzp}.

Recently the interest to extended MHD effects had been renewed due to peculiar behaviour of Z-pinch experiments 
with the axial imposed magnetic field. In these experiments the mirror symmetry of classic Z-pinches is explicitly broken by the axial field and the Hall term can become very important in the low-density peripheral plasma, dramatically changing current distribution and azimuthal field, presumably due to helical instabilities\cite{mikitchuk2019}. Similar effect have been also observed in extended-MHD simulations with axial field\cite{seyler2018,seyler2020}. It remains to be seen if such effects can be observed when mirror symmetry is broken by the geometry of the experiment rather than axial magnetic field.

The data that support the findings of this study are available from the corresponding author upon reasonable request.
The code used to generate most of the data is openly available\cite{verif_coupling}.

\section{Acknowledgement}
We are grateful to Dr. C.~L.~Rousculp, Dr. S. T. Zalesak, Prof. C. E. Seyler, and Prof. A. B. Sefkow for insightful comments.
A. Beresnyak, A. Velikovich, J. Giuliani and A. Dasgupta were supported by the Department of Energy/National Nuclear Security Administration under the Interagency Agreement DE-NA0003278. A. Beresnyak also had support from DOE/NNSA ASC program. 
S. L. Jackson, J. T. Engelbrecht and A. S. Richardson and A. Beresnyak were supported by the Naval Research Laboratory through the Basic and Applied Research Programs.

\par
\ \ 
\par

\def\apj{{\rm Astrophys. J.}}           
\def\apjl{{\rm ApJ }}          
\def\apjs{{\rm ApJ }}          
\def\grl{{\rm GRL }}
\def\aap{{\rm A\&A } }
\def\jgr{{\rm JGR}}
\def\mnras{{\rm MNRAS } }
\def\physrep{{\rm Phys. Rep. } }               
\def\prl{{\rm Phys. Rev. Lett.}} 
\def\pre{{\rm Phys. Rev. E}} 
\def\araa{{\rm Ann. Rep. A\&A } }
\def\prd{{\rm Phys. Rev. D}} 
\def\pra{{\rm Phys. Rev. A}} 
\def\ssr{{\rm SSR}}
\def\planss{{\rm Plan. Space Sciences}}
\def\apss{{\rm Astrophysics and Space Science}}
\def\nat{{\rm Nature}}
\def\jcap{{\rm JCAP}}
\def\memsai{{\rm MEMSAI}}
\def\lt{{<}}
\def\aapr{{\rm AAPR}}
\def\solphys{{\rm SolPhys}}

\bibliography{dpf}

\end{document}